\documentclass[12pt]{article}
\usepackage{fullpage}
\usepackage{natbib}

\title{Heuristic Representations of Plasma Momentum Transport}
\author{I.H.Hutchinson,\\ Plasma Science and Fusion Center
  and\\ Department of Nuclear Science and Engineering\\ Masschusetts
  Institute of Technology,\\ Cambridge, MA, USA}

\begin{document}

\maketitle

\begin{abstract}
It is argued that the form commonly adopted for heuristic
representation of tokamak momentum diffusion has major shortcomings, and that
a non-diffusive momentum-flux term independent of velocity is more
appropriate. 
\end{abstract}

\section{Introduction}

In attempting to understand the self-acceleration of tokamak plasmas\cite{hutchinson00}
and other intriguing momentum transport phenomena in magnetically
confined plasmas, a heuristic model of plasma momentum transport has
frequently been adopted which separates the momentum transport flux
into ``diffusive'' and ``convective''
parts\cite{rice04,peeters07,gurcan08,eriksson09}. It is shown here
that, in this division, which in the absence of fundamental
calculations of the transport is \emph{ ad hoc}, the convective part
cannot consistently be identified with a term that is equal to the
plasma momentum density times a constant coefficient, as is often
implied or assumed. Instead, at least a part, and arguably the most
interesting and significant part, of the momentum flux must be
represented as a term independent of the velocity, rather than
proportional to velocity. This velocity-independent part of the
momentum flux does not represent, in any meaningful sense, a momentum
source. It represents a transport flux of momentum up the velocity
gradient that is independent of both the velocity gradient itself
(i.e.\ it is not a form of viscosity) and of the value of the
velocity. The latter property is characteristic of momentum that is
transported not by mean particle flux but by distinctively momentum
transport processes such as by Reynolds stress. Experiments, if they
are sufficiently comprehensive, could in principle identify the
velocity dependence, if any, of the total ``convective'' momentum
transport, but in practice this is beyond the measurement capability
so far.

\section{Heuristic flux density representations}

\subsection{Fick's Law and its failure}
A familiar description of diffusing systems is ``Fick's Law'' which
assumes that diffusive flux density $\Gamma_Q$ is proportional to the gradient
of the property under consideration which we will denote here $Q$.
\begin{equation}\label{Fick}
  \Gamma_Q = D_Q {dQ \over dx}\ ,
\end{equation}
where $D_Q$ is called the diffusivity or diffusion coefficient.
For convenience we are using a one-dimensional, slab representation in
this discussion which is sufficient for immediate purposes.

For treatment of most plasma magnetic confinement problems, this
simple approximation is known to be inadequate. Fluxes are present
that prove \emph{not} to be proportional to the gradient of the
extensive quantity $Q$. Consequently, while it is still possible to
write a Fick's law and consider it to be the definition of the
diffusion coefficient $D_Q$, that diffusion coefficient $D_Q$ is not
constant, and not even necessarily non-singular (where $dQ/dx = 0$ for
example $D_Q$ becomes formally infinite if the flux there is
non-zero).  Therefore, for a consistent representation, at a minimum
an additional term must be included in the expression for the momentum
flux. The extra term that is added to improve the heuristic transport
representation ought generally to avoid the singularity in $D_Q$
without introducing singularities of its own. Ideally it would
represent an identifiable physical process, but that is not
guaranteed by the formal representation.

\subsection{Density diffusion}

When the quantity under discussion is the
density, it is satisfactory to express the additional flux term in the
form $V_Q Q$, because $n$ never changes sign, and so singularities in
$V_Q$, which is called the convection velocity, are not induced. In
short, it makes reasonable physical sense to write
\begin{equation}\label{density}
  \Gamma_n = D_n {dn\over dx} + V_n n \ .
\end{equation}
And indeed the ``diffusion'' (first) and ``convection'' (second) terms
have some mathematical and physical justification in terms of fluid
transport.  [But notice that the convection velocity, $V_n$, is not in
  general equal to the fluid velocity $\Gamma_n/n$.]

Let us note, however, that eq (\ref{density}) (like eq \ref{Fick}) is a purely
formal expression. Any flux density could be written this way, as a
pure piece of arithmetic, with the only constraint on the $D_n$ and
$V_n$ being that the diffusion term and the convection term add up to
the correct total flux. The form is suggestive of diagonal and
off-diagonal terms in a transport matrix. If such a representation were
valid, then one might find that $D_n$ and $V_n$ were independent of the
local density and its gradient. Only in the case of approximately
constant $D_n$ and $V_n$ does the division into these two terms really have any significance.

\subsection{Momentum transport}

When, instead, the quantity being transported is the momentum density
in a direction ($z$) transverse to the direction of transport, $n m
v_z$, it is not at all obvious that the second term which seeks to
generalize Fick's law, should be written proportional to the momentum
density $nmv_z$. The first obvious fact is that $v_z$ can have either
sign; and that there are therefore likely places where it goes through
zero. Unless the additional momentum flux term is identically zero
there, a singularity in the coefficient multiplying $nmv_z$ will
arise. Of course, the arithmetic might be fixed up by adjusting $D_v$ so
that the non-diffusive flux is in fact always zero where $v_z$ is
zero, but such an adjustment cannot be done with constant
$D_v$. Therefore writing the momentum flux, as several recent authors
have done, as
\begin{equation}\label{badeq}
  \Gamma_v = D_v{d\over dx} (nmv_z) + V_c nmv_z,  \ {\rm or} \quad \Gamma_v= nm\left(D_v{d\over dx} v_z + V_c v_z\right)
\end{equation}
has little to recommend it. It introduces by a pure mathematical
ansatz a velocity $V_c$ that will generally have to become infinite
where $v_z$ changes sign. To avoid such infinities one would have to
do fine tuning of $D_v$, which contradicts the observation made in the
previous section that the division into $D_v$ and $V_c$ has significance
only when specific constraints, such as that $D_v$ and $V_c$ are
invariant, are applied.

Worse still, the form of eq (\ref{badeq}) violates a fundamental
property that we should require of non-relativistic physical equations
in a translationally-invariant configuration: that they be invariant
under Galilean transformation. If we change our frame of reference to
one moving with velocity $v_t$ in the $\hat{z}$-direction, then all
transformed (primed) velocities are related to the untransformed-frame
velocities via $v_z' = v_z - v_t$. Eq (\ref{badeq}) (in its second
form for simplicity) becomes
\begin{equation}\label{transformed}
  \Gamma_v'= nm\left(D_v{d\over dx} v_z' + V_c v_z' + V_c v_t\right)\ ,
\end{equation}
yielding a flux in the $\hat{x}$-direction that is \emph{not}
invariant under the transformation, even though, physically, it should
be, and indeed yielding a form of the equations in which a new term,
$V_cv_t$, independent of both $dv'/dx$ and $v'$ has entered. Expanding
on this point in the context of momentum transport, one should realize
that if there is a preferred frame of reference, the laboratory frame,
for the solution, then that preference is imposed not via the
transport equations themselves, but by the boundary condition. In
other words, for a translationally invariant confined plasma, there is
nothing in the transport \emph{equations} that specifies the
laboratory frame, only in the \emph{boundary conditions} such as that
the velocity $v_z$ should be zero at the walls in a no-slip
situation. [If there were \emph{momentum sources} these might also
  select a preferred frame, but the point remains that the
  flux-density itself should not].

The whole spirit of the heuristic diffusive representation
presumes this view. It is therefore a major shortcoming of any
proposed heuristic transport expression that it not exhibit Galilean
invariance with respect to velocity in a direction of
symmetry. Supposing the non-diffusive part of momentum transport to be
$V_c nmv_z$ as, eq (\ref{badeq}) does, has precisely this shortcoming.

This invariance failure is partially obscured in a toroidal situation,
because transformation to a \emph{rotating} frame of reference itself
introduces new terms into the plasma equations: the centrifugal and
coriolis forces\cite{peeters07}. Since the centrifugal force is an
even (and second order) function of the toroidal rotation velocity, it
can hardly be a source of rotation. But the coriolis force is not so
obviously irrelevant. Nevertheless, it seems unlikely that its effect
can explain self-acceleration; not least because the coriolis force
does not of itself show a systematic preferred sign of rotation (which
the experiments do); it merely amplifies rotation with one or the
other sign.  It is also certain that constraining the non-diffusive
momentum transport term to have a form that is exactly consistent with
arising from coriolis forces is, \emph{a priori}, unjustified.

\subsection{Momentum convection}

There is, in fact, a part of the momentum flux-density that is truly
convective and ought to be written $nmv_z v_x = \Gamma_n mv_z$. It is
the actual momentum convection arising from the plasma flow velocity
$v_x=\Gamma_n/n$. This elementary term is of course present in any
fluid representation, and is the flux of momentum carried by the
(total) particle flux density $\Gamma_n$. This term, while potentially
important in transients and essential for maintaining the Galilean
invariance in the presence of particle flux, is not of the slightest
value as an explanation for tokamak self-acceleration, because that
acceleration occurs in plasmas with zero particle source and steady
density profiles, for which $\Gamma_n=0$. Therefore, while one might
call the momentum flux-density that arises from a particle pinch such
as the Ware pinch a ``momentum pinch'', it is completely irrelevant to
the explanation of self-acceleration. A particle pinch can plausibly
be invoked to explain non-uniform density profiles when there is zero
density source and yet $D_n\not=0$. But in the interesting steady
situations, such a particle pinch is always being exactly cancelled by
density diffusion, yielding $\Gamma_n=0$ and hence zero momentum
convection.

For example, addressing specifically the turbulent transport
situation, fluxes of particles and momentum can be considered to arise
from the averages of the turbulent transport terms as
$\Gamma_n=\langle\tilde v_x \tilde n\rangle$ and $\Gamma_v=m\langle
\tilde v_x \widetilde{n v}_z\rangle = m\langle v_z\rangle\langle\tilde
v_x\tilde n\rangle + m\langle n\rangle\langle\tilde v_x \tilde
v_z\rangle = \Gamma_n mv_z + m \langle n\rangle \langle\tilde v_x
\tilde v_z\rangle$, where the last term is directly identifiable as
the Reynolds stress, and is manifestly independent of (Galilean) frame
of reference.

\subsection{Self-consistent heuristic momentum flux density}

In view of the foregoing considerations, the minimum self-consistent
heuristic representation of momentum flux density useful to describe a
combination of diffusive and non-diffusive momentum transport is of the form:
\begin{equation}\label{diffconv}
  \Gamma_v = nm\left(D_v {dv_z\over dx} + {\Gamma_n\over n} v_z + V_v
  v_0  \right) ,
\end{equation}
where $v_0$ is merely a conveniently chosen constant characteristic
velocity (e.g. the sound speed) that renders the term $V_vv_0$
dimensionally as a product of two velocities, independent of
$v_z$. Only their product, $V_vv_0$, is significant. The middle term,
proportional to $v_z$, must be chosen so as to annihilate the particle
density divergence effects arising in Galilean transformations; so its
coefficient should not be regarded as a free parameter, but is
$\Gamma_n/n$. This term is not relevant to understanding self-acceleration.

\section{Sources, conservation equations and their inversion}

\subsection{Conservation equation structure}

The conservation of plasma $\hat{z}$-momentum in a cylindrically symmetric
system is 
\begin{equation}\label{momcons}
  {\partial\over\partial t}(nmv_z) = \nabla. \Gamma_v + S_v =
  {\partial\over r\partial r}( r\Gamma_{vr}) + S_v ,
\end{equation}
where $S_v$ is a possible local internal momentum source density. The
analytic boundary condition on axis is $\Gamma_{vr}(0)=0$; therefore
without internal sources ($S_v=0$), in steady state
($\partial/\partial t=0$), the solution is $\Gamma_{vr}=0$. If
$\Gamma_v$ is of the form (\ref{diffconv}), then trivially
\begin{equation}
  {dv_z\over dr} + {\Gamma_n\over D_v n} v_z + {V_v\over D_v} v_o =0\ .
\end{equation}
If the density is also governed by a steady sourceless diffusion equation,
then $\Gamma_n=0$ and we find
\begin{equation}
  v_z = v_{za} -  \int_a^r {V_v\over D_v} v_o dr\ ,
\end{equation}
where the edge ($r=a$) boundary condition is $v_z=v_{za}$.

Notice that for this source-free situation the term $V_vv_0$ is the
cause of any non-zero velocity gradient. The role it serves is very
similar to that of the source, $S_v$, in a situation where the
flux-density is purely diffusive, $\Gamma_v=D_v nm dv/dr$. In that
pure-diffusive case, without sources the steady solution for $v_z$ is uniform,
equal to the edge value. But we should not be misled into calling $n m
v_0V_v$ (or more precisely its divergence) a momentum
\emph{source}. It is not a source; it is a momentum flux density that
occurs independent of velocity gradient.

More generally, substituting the form (\ref{diffconv}) into eq
(\ref{momcons}), we obtain
\begin{equation}
  {\partial\over\partial t}(nmv_z) - S_v = {\partial\over r\partial r}\left[ r 
nm\left(D_v {\partial v_z\over \partial r} + {\Gamma_n\over n} v_z + V_v
  v_0  \right)\right].
\end{equation}
For simplicity of discussion of the velocity, let's assume that
density $n$ is fixed and uniform (general cases have more terms to
consider) and that $\Gamma_n=0$, so that this
can be written:
\begin{equation}\label{Leq}
    L\equiv {\partial\over\partial t}(v_z) - S_v/nm  = {\partial\over
      r\partial r}\left[ r \left(D_v {\partial v_z\over \partial r} + V_v v_0
      \right)\right]
\end{equation}
\begin{equation}
={1\over r}\left[{\partial\over
      \partial r}\left(r D_v\right) {\partial v_z\over \partial r} + (r D_v) {\partial\over
      \partial r}  {\partial v_z\over \partial r} + v_o{\partial\over
      \partial r} \left(rV_v\right)\right].
  \end{equation}

\subsection{Deducing $D$ and $V$ from measurements}

If we wish to deduce from measurements the values of $D_v$ and
$V_v$, then the left-hand-side terms of eq (\ref{Leq}) act in essentially
the same way. There are either transients ($\partial v_z/\partial t$)
or sources ($S_v$). If neither is present, then the equations are
homogeneous, requiring simply $D_v {\partial v_z/\partial r} + V_v v_0=0$,
and showing that there is nothing setting the overall size of $D$ or
$V$, only their ratio $D_v/V_v$. When non-zero, the LHS determines the
divergence of the flux (the total of the RHS), and acts as an
inhomogeneous contribution to the radial equation governing the
quantities $rD_v$, and $r V_v$.

If one has measurements of ${\partial v_z/\partial r}$ for all
relevant radii, then instantaneously eq (\ref{Leq}) is simple linear functional
equation for $(rD_v), (rV_v)$. But its solution is not unique. For
example one possible solution is $D_v=0$ and $\partial rV_v/\partial r
= rL/v_0$, another is to take $V_v=0$.  On the basis of only one
instantaneous (or steady) case nothing can be done to narrow down the
possible solution space. In other words, $D$ and $V$ cannot be separated.
However, if a range of different profiles of $L(r)$ is available, for
example because a transient can be time-resolved to give a variety of
different $L$, or because the
sources' spatial profile can be varied, and if ${\partial v_z/\partial
  r}$ is measured for each case, then some best-fit for $rD_v$ and
$rV_v$ can be determined (as a function of $r$), \emph{provided they
  are the same for all cases}. Without this last stipulation (or some
other about how the $D$ and $V$ for different cases are related), then
one is no better off than with one instantaneous case.

From a practical viewpoint, the process of solving for $rD_v$ and
$rV_v$ will presumably be to (1) discretize them into a finite
representation in terms of a set of coefficients times appropriate
functions, (2) obtain the linear (matrix) relationships between the
coefficients of the discretized representation and the values of
$L(r,t)$, (3) invert the matrix equation, probably in some
regularized least-squares sense, to obtain the coefficients. If, as is
usually the case, the amount of linearly independent information is
strongly limited by constraints on the experiments and on their
uncertainty, then a very judicious choice of representation is
important, and only very few coefficients can be deduced. For example,
perhaps the simplest low order representation is to take $D_v=const$
(independent of $r$), and $V_v= rV_{va}/a$, just two independent
coefficients ($D$ and $V_{va}/a$). The $V_v$ representation has been
chosen recognizing that analyticity on axis requires $V_v=0$ there.
Then in principle, two independent profiles of $L$ provide sufficient
rank to solve for the coefficients. Of course this is no guarantee
that the profiles chosen are correct, and preferably a much bigger
range of profiles and more comprehensive functional representation
should be used. 

\subsection{Identifying flow-drive}

In particular, in flow-drive experiments (e.g.\cite{lin09}), it would be
completely unjustified to assume a highly simplified representation of
the form $D_v=const$ and $V_v= rV_{va}/a$ and then try to invert the
process and deduce from the profiles of $v_z$ the spatial dependence
of $S_v$. An identification of unknown source profiles would instead
have to rely for example upon the RHS of eq (\ref{Leq}), the transport
terms, being negligible compared with inertia for sufficiently rapid
accelerations, in a manner analogous to the ``break of slope''
analysis of heating steps. Present experiments don't normally have
such a clear separation of timescales for momentum source
transients. The result is that there is no basis on which to
distinguish between changes in the term $v_o \partial(rV_v)/\partial
r$, and changes in the term $S_v/nm$. In other words, one can't tell
from velocity measurements whether there is actually a momentum source
($S_v$) or whether the flow-drive works by changing the momentum
transport ($V_v$). In the latter case, if the effect on $V_v$ is
localized in radius, the radial integral of the $v_o
\partial(rV_v)/\partial r$ term across the whole perturbed region is
zero: an embodiment of the fact that the total equivalent force
arising from this term is zero. Sometimes this is summarized by
referring to the transport-alteration influence as ``dipolar''. If the
transport alteration is zero outside some radius $r_o$, and there are
no changes to $S_v$, then the solution of the conservation equation
should give unperturbed velocity profile outside $r_0$.  But if the
transport alteration extends all the way to the boundary, so that
$V_v$ is altered at the boundary, this conservation no longer applies,
the influence need not be ``dipolar'', and the velocity perturbation
can extend to the edge.

It should be obvious from the above that knowing how the $V_vv_o$ term
varies with $v_z$ is crucial. Here, we've taken it to be independent
of $v_z$ because of the arguments about Galilean invariance. In
principle we could include two different terms $V_vv_o$ and $V_cv_z$,
and then solve for each, if we have sufficient linearly independent
$L$ values, thus identifying experimentally the balance between the
two types of term. But it is already a stretch for most experiments to
separate $D$ from $V$ at all, let alone bringing in other possible terms.
Tokamak experiments so far have not had the precision or comprehensive
resolution to do so. 

\bibliography{momdiff}

\begin{thebibliography}{1}

\bibitem{hutchinson00}
I~H Hutchinson, J~E Rice, R~S Granetz, and Snipes.
\newblock Self-acceleration of a tokamak plasma during ohmic h-mode.
\newblock {\em Physical Review Letters}, 84:3330, 2000.

\bibitem{rice04}
J.E. Rice, W.D. Lee, E.S. Marmar, P.T. Bonoli, R.S. Granetz, M.J. Greenwald,
  A.E. Hubbard, I.H. Hutchinson, J.H. Irby, Y.~Lin, D.~Mossessian, J.A. Snipes,
  S.M. Wolfe, and S.J. Wukitch.
\newblock Observations of anomalous momentum transport in alcator c-mod plasmas
  with no momentum input.
\newblock {\em Nuclear Fusion}, 44(3):379--386, 2004.

\bibitem{peeters07}
A.~G. Peeters, C.~Angioni, and D.~Strintzi.
\newblock Toroidal momentum pinch velocity due to the coriolis drift effect on
  small scale instabilities in a toroidal plasma.
\newblock {\em Physical Review Letters}, 98:265003, 2007.

\bibitem{gurcan08}
O.~D. Gurcan, P.~H. Diamond, and T.~S. Hahm.
\newblock Turbulent equipartition and homogenization of plasma angular
  momentum.
\newblock {\em Physical Review Letters}, 100(13):135001, 2008.

\bibitem{eriksson09}
L-G Eriksson, T~Hellsten, M~F~F Nave, J~Brzozowski, K~Holmstrom, T~Johnson,
  J~Ongena, K-D Zastrow, and JET-EFDA Contributors.
\newblock Toroidal rotation in rf heated jet plasmas.
\newblock {\em Plasma Physics and Controlled Fusion}, 51(4):044008 (23pp),
  2009.

\bibitem{lin09}
Y~Lin, J~E Rice, S~J Wukitch, M~J Greenwald, A~E Hubbard, A~Ince-Cushman,
  L~Lin, E~S Marmar, M~Porkolab, M~L Reinke, N~Tsujii, J~C Wright, and Alcator
  C-Mod Team.
\newblock Observation of ion cyclotron range of frequencies mode conversion
  plasma flow drive on alcator c-mod.
\newblock {\em Physics of Plasmas}, 11:056102, 2009.

\end{thebibliography}

\end{document}